\documentclass[12pt]{iopart}
\usepackage[dvips]{graphicx}
\usepackage{graphicx}
\usepackage{latexsym}
\graphicspath{{plots/}}
\DeclareGraphicsExtensions{.eps.gz,.eps,.ps,.ps.gz}
\newcommand{\lwig}{\mbox{\,\raisebox{.3ex}
    {$<$}$\!\!\!\!\!$\raisebox{-.9ex}{$\sim$}\,}}
\newcommand{\gwig}{\mbox{\,\raisebox{.3ex}
    {$>$}$\!\!\!\!\!$\raisebox{-.9ex}{$\sim$}}\,}

\newcommand{\xpr}{{x^\prime}}
\newcommand{\ii}{{\rm i}}

\newcommand{\iai}{I\overline{I}} 

\def\Journal#1#2#3#4{#4 {#1} {#2} #3}
\def\CPC{\em Comput. Phys. Commun.}

\def\EPJ{\em Eur. Phys. J.}
\def\PREP{\em Phys. Rep.}
\def\MPL{\em Mod. Phys. Lett.}
\def\TMP{\em Theor. Math. Phys.} 
\def\funp{{I\!\!P}}

\begin{document}
{\normalsize\rightline{DESY 01-125}}
\title{Tracking QCD--Instantons}  
\author{F. Schrempp}
\address{Deutsches Elektronen-Synchrotron DESY, Hamburg, Germany} 
\begin{abstract}
In a first part, I review our detailed investigation of the prospects
to discover deep-inelastic processes induced by small QCD-instantons
at HERA. This includes the essence of our calculations based on
instanton-perturbation theory, crucial lattice  constraints 
and a confrontation of the recent intriguing  (preliminary) search results
by the H1 collaboration with our predictions. In a second part, I report  
on two ongoing attempts towards a better understanding of the r{\^o}le of
larger-size instantons in the QCD-vacuum and in high-energy
(diffractive) scattering
processes, respectively: The striking suppression of larger-size
instantons in the vacuum is attributed to a residual conformal
inversion symmetry, and a precocious lack of ``color
transparency'' in the one-instanton contribution to the color-dipole scattering
picture is emphasized.  
\end{abstract}
\section{Introduction}
Instantons\cite{bpst,th} represent a basic non-perturbative aspect of QCD,
but their direct experimental verification is still lacking. Being
topologically non-trivial fluctuations of the gluon fields,
QCD-instantons induce hard, chirality-violating  processes absent in 
conventional perturbation theory\cite{th}. Deep-inelastic scattering
(DIS) at HERA has been shown to offer a unique opportunity\cite{rs1}
to discover such processes induced by {\it small} instantons ($I$)
through a sizable\footnote[1]{For an exploratory 
calculation of the instanton contribution to the gluon-structure
function, see Ref.~\cite{bb}.} rate\cite{mrs,rs2,rs-lat}
and a characteristic final-state signature~\cite{rs1,cgrs,qcdins,rs3}. 
A first part of this talk is devoted to a review of our extensive
investigation on small instantons in deep-inelastic scattering. This
includes a ``flow-chart'' of our calculations based on
$I$-perturbation theory\cite{mrs,rs2}, an exploitation of crucial lattice
constraints\cite{rs-lat,ukqcd} and a 
confrontation\cite{rs3} of the recent intriguing (preliminary) $I$-search
results by the H1 collaboration\cite{mikocki,h1_ichep,koblitz} with
our predictions.  
In a second part, I report on two ongoing
attempts\cite{conformal,dipole-ins} towards a better 
understanding of the r{\^o}le of {\it larger-size} instantons in the
QCD-vacuum and in high-energy (diffractive) scattering 
processes, respectively: The striking suppression of larger-size
instantons in the vacuum is attributed to a residual {\it conformal
inversion} symmetry\cite{conformal}, and an interesting,
precocious lack of ``color transparency'' in the one-instanton
contribution\cite{dipole-ins} to the 
color-dipole scattering picture\cite{dipole} in the deep-inelastic
regime  is emphasized. 
\section{Small instantons and deep-inelastic scattering}
\subsection{Instanton-perturbation theory}
Let us start by briefly summarizing the essence of
our theoretical calculations\cite{mrs,rs2} based on so-called
$I$-perturbation theory. As we shall see below, in an appropriate
phase-space region of deep-inelastic scattering with generic hard
scale $\mathcal{Q}$, 
the contributing $I$'s and $\overline{I}$'s have {\it small size}
$\rho\lwig\mathcal{O}(\frac{1}{\alpha_s(\mathcal{Q})\mathcal{Q}})$ and may be self-consistently considered as a
{\it dilute} gas, with the small QCD coupling $\alpha_s(\mathcal{Q})$ being the
expansion parameter like in usual perturbative QCD (pQCD). Unlike the
familiar expansion about 
the trivial vacuum $A^{(0)}_\mu=0$ in pQCD, in $I$-perturbation theory   
the path integral for the generating functional of the   
Green's functions in Euclidean position space is then expanded about the
known, classical one-instanton solution, $A_\mu=A^{(I)}_\mu(x)+\ldots$.
After Fourier transformation to momentum space, LSZ amputation and
careful analytic continuation to Minkowski space (where the actual
on-shell limits are taken), one obtains a corresponding set of modified Feynman
rules for calculating $I$-induced scattering amplitudes.  
As a further prerequisite, the masses $m_q$ of the active quark
flavours must be light on the scale of the inverse effective $I$-size
$1/\rho_{\rm eff}$, i.\,e.
$ {\rm m}_q\cdot\rho_{\rm eff}\ll 1$.

\begin{figure}
\begin{center}
\parbox{10cm}{\includegraphics*[width=10cm]{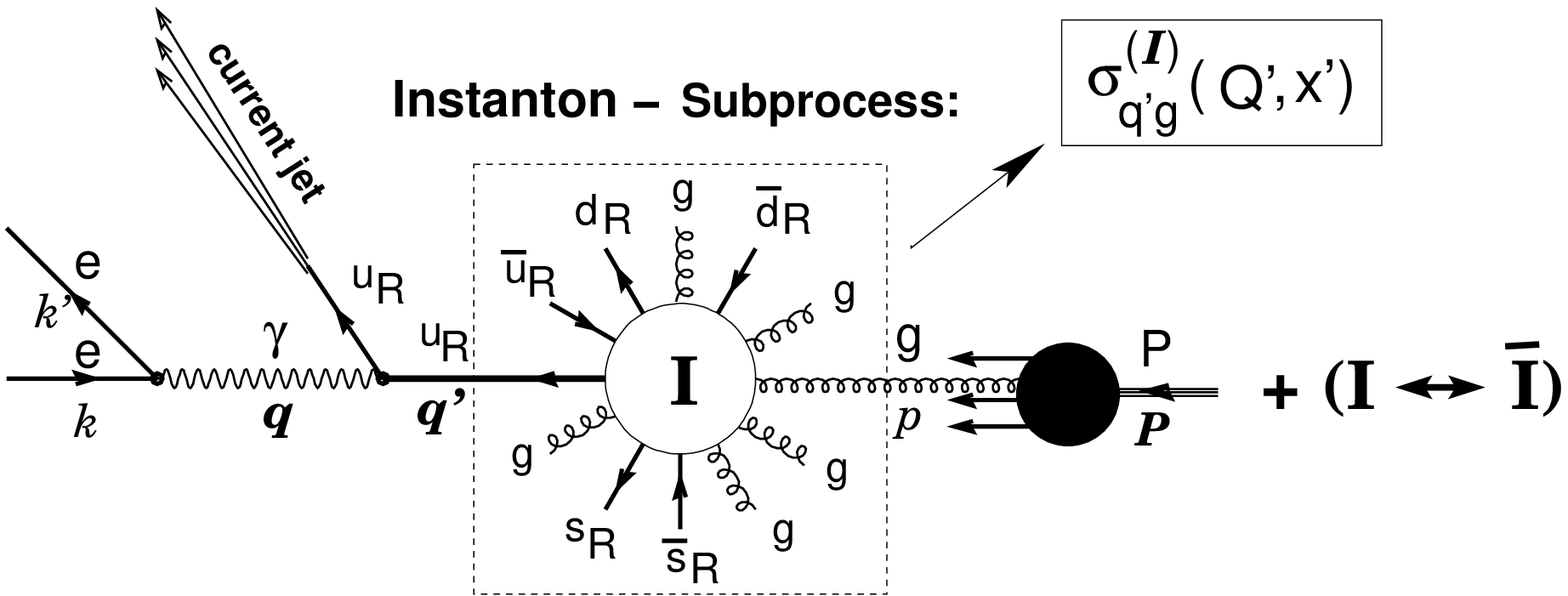}}\hfill
\parbox{5cm}{\includegraphics*[width=5cm]{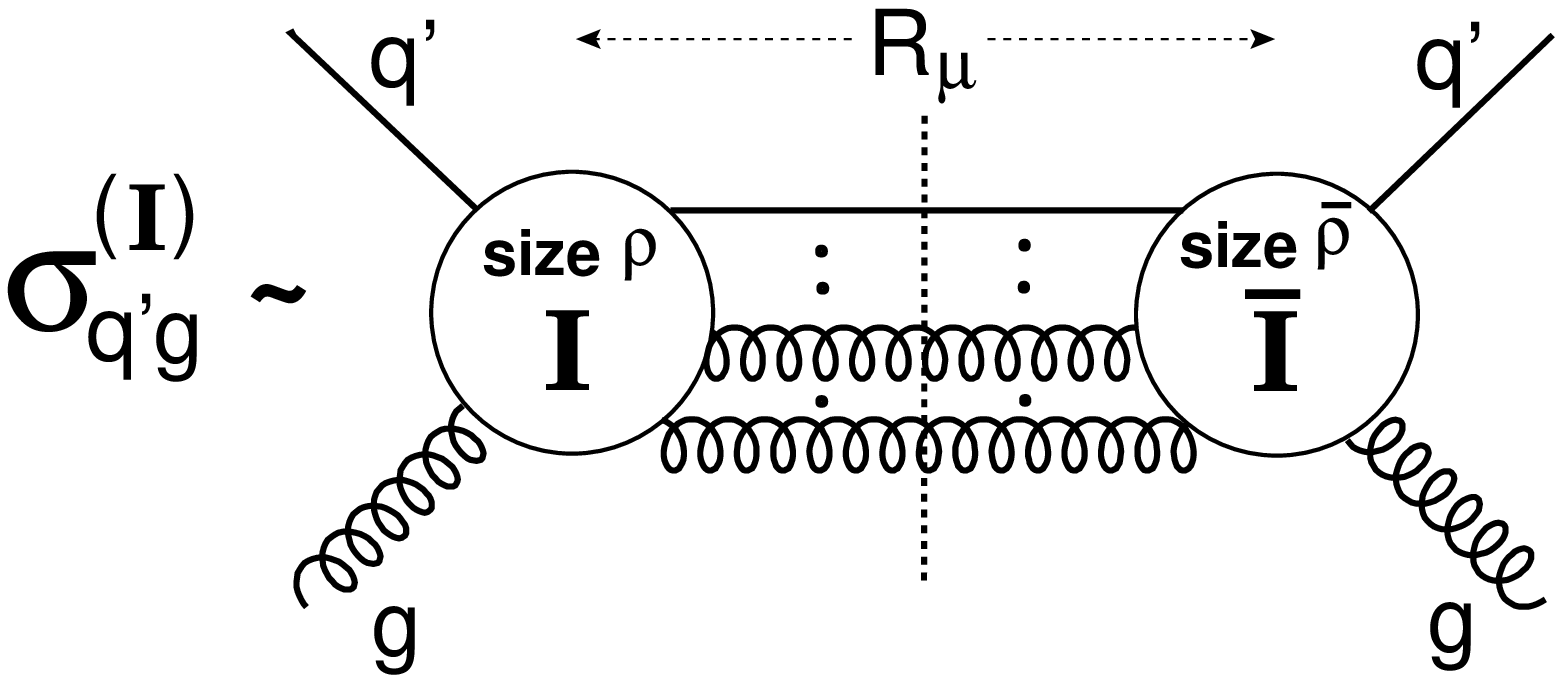}}
\caption[dum]{\label{lead} (left): Leading, instanton-induced 
process in deep-inelastic $e^\pm {\rm P}$ scattering for
$n_f=3$. (right): Structure of the total cross section for the
chirality-violating ``instanton-subprocess'' $q^\prime\,g
\stackrel{(I)}{\Rightarrow} X$ and illustration of the collective
coordinates $\rho,\overline{\rho},R_\mu$.   
 }
\end{center}
\end{figure}
The leading, $I$-induced, chirality-violating
process in the deep-inelastic regime of $e^\pm {\rm P}$ scattering is
displayed in \Fref{lead}\,(left) for $n_f=3$ $~$ massless flavors. In the
background of an $I\ (\overline{I}\,)$ (of topological charge $Q=+1\
(-1)$), all $n_f$ massless quarks and anti-quarks are right
(left)-handed such that the $I$-induced 
subprocess emphasized in \Fref{lead}\,(left) involves a violation of
chirality $Q_5=\#\,(q_{\rm R}+\overline{q}_{\rm R})-\#\,(q_{\rm
L}+\overline{q}_{\rm L})$ by an amount,  
\begin{equation}
\Delta Q_5 =2\,n_f\,Q, 
\end{equation}
in accord with the general chiral anomaly relation\cite{th}.  
Within $I$-perturbation theory, one first of all derives
the following factorized expression in the Bjorken limit of
the $I$-subprocess variables 
$Q^{\prime\,2}$ and $\xpr$ (c.\,f. \Fref{lead}\,(left)): 
\begin{equation}
\frac{{\rm d}\sigma_{\rm  HERA}^{({ I})}}{{\rm d}{\xpr} {\rm
d}{ Q^{\prime\,2}}}\simeq \frac{ {\rm d}{\cal
L}_{{ q^\prime} g}^{({ I})}}{{\rm d}{\xpr} {\rm
d}{ Q^{\prime\,2}}}\cdot { 
\sigma_{{ q^\prime} g}^{({ I})}({ Q^\prime},{\xpr})}
\hspace{3ex}{\rm for\ }\left\{ 
\begin{array}{l} 
Q^{\prime\,2}=-q^{\prime\,2}>0 { \rm \ large},\\  
0\le \xpr=\frac{Q^{\prime\,2}}{2 p\cdot q^\prime}\le 1 {\rm \ fixed\ }.
\end{array} 
\right .
\label{Bjlim}
\end{equation}
In \Eref{Bjlim}, the differential luminosity, ${\rm d}{\cal
L}_{q^\prime\,g}^{( I)}$ counts the number of $q^\prime\,  g$
collisions per $eP$ collisions. It is  given in terms of integrals over
the gluon density, the virtual photon flux, and the (known) flux of
the virtual quark $q^\prime$ in the instanton background\cite{rs2}. 

The essential instanton dynamics resides, however,  in the total
cross-section of the $I$-subprocess $q^\prime\ g
\stackrel{I}{\Rightarrow} X$ as emphasized in \Fref{lead}. 
Being an observable, $\sigma_{q^\prime g}^{(I)}( Q^\prime, \xpr)$ involves 
integrations over all $I$ and $\overline{I}\,$-``collective coordinates'',
i.\,e. the $I\ (\overline{I}\,)$ sizes  $\rho\ (\overline{\rho}\,)$, 
the $\iai$ distance four-vector  $R_\mu$ and the relative $\iai$ color
orientation matrix $U$.  

\begin{equation}
\fl \sigma^{(I)}_{{  q^\prime}\,g}=
      \int d^4 { R}\ 
      {\rm e}^{\ii\, (p+{ 
      q^\prime})\cdot { R}} 
      \int\limits_0^\infty d{ \rho}
      \int\limits_0^\infty d{ \overline{\rho}}\
 {{\rm e}^{-({\rho +\overline{\rho}}){ Q^\prime}}\,}
 \ {
        D({ \rho})\, D(\overline{\rho}\,)}
      \int d{ U}\,{\rm e}^{{-\frac{4\pi}{\alpha_s}}\,
      { \Omega\left({
      U},{\frac{ R^2}{\rho\overline{\rho}}},{ 
      \frac{\overline{\rho}}{\rho}} \right)}}\,\{\ldots\}\,
\label{cs}
\end{equation}
Both instanton and anti-instanton degrees of freedom enter here, 
since cross-sections result from taking the modulus squared of an
amplitude in the single $I$-background. Alternatively and more conveniently
(c.\,f. \Fref{lead}\,(right)), one may obtain the cross-section
(\ref{cs}) as a discontinuity of the $q^\prime\,  
g$ forward elastic scattering amplitude in the
$\iai$-background~\cite{rs2}. The $\{\ldots\}$ in \Eref{cs} abreviates
smooth contributions associated with the external partons etc.
   
Let us concentrate on two crucial and strongly varying quantities of the
$I$-calculus appearing in \Eref{cs}: $D(\rho)$, the (reduced) $I$-size
distribution\cite{th,bernard}, and
$\Omega\left(U,\frac{R^2}{\rho\overline{\rho}}, 
 \frac{\overline{\rho}}{\rho} \right)$, the $\iai$
 interaction\cite{valley1,valley2}, associated with a resummation of
 final-state gluons. Both objects are {\it known} within
 $I$-perturbation theory, 
 formally for  $\alpha_s(\mu_r)\ln(\mu_r\,\rho)\ll 1$ and 
$\frac{R^2}{\rho\overline{\rho}} \gg 1$ (diluteness), respectively,
with $\mu_r$ being the renormalization scale. 

Most importantly, the resulting power-law behaviour for the $I$-size
distribution,
\begin{equation}
D(\rho)\propto \rho^{\beta_0-5 +\mathcal{O}(\alpha_s)}, 
\end{equation} 
involving the leading QCD $\beta$-function coefficient,
$\beta_0=\frac{11}{3}\,N_c-\frac{2}{3}\,n_f,\ (N_c=3)$, 
generically spoils the calculability of $I$-observables due to the
bad IR-divergence of the integrations over the $I\
(\overline{I}\,)$-sizes   for large $\rho\ (\overline{\rho}\,)\,$. 
Deep-inelastic scattering represents, however, a crucial exception: 
The {\it exponential} ``form factor''
$\exp({- Q^\prime}(\rho+\overline{\rho}\,))$ that was shown\cite{mrs}
to arise in \Eref{cs}, insures convergence and {\it small} instantons
for large enough $Q^\prime$, despite the strong power-law growth of $D(\rho)$. 
This is the key feature, warranting the calculability of
$I$-predictions for DIS.

It turns out that for (large) $Q^\prime \ne 0$, all collective
coordinate integrations in $\sigma_{q^\prime g}^{(I)}$ of \Eref{cs} may be
performed in terms of a {\it unique saddle point}: 
\begin{eqnarray}
U^\ast&\Leftrightarrow& \mbox{\rm \ most attractive relative $\iai$
orientation in color space},\nonumber\\ 
\rho^\ast &=& \overline{\rho}^\ast\sim
\frac{4\pi}{\alpha_s(\frac{1}{\rho^\ast})}\,\frac{1}{ Q^\prime};    
\hspace{2ex}
\frac{
 R^{\ast\,2}}{\rho^{\ast\,2}} \stackrel{ \ Q^\prime {\rm\ large}\
 }{{\sim}} 4\frac{\xpr}{ 1-\xpr}   
\label{saddle}
\end{eqnarray}
This result underligns the self-consistency of the approach, since for
large $Q^\prime$ and  small $(1-\xpr)$ the  
saddle point (\ref{saddle}), indeed, corresponds to widely separated, small
$I$'s and $\overline{I}$'s.  

\subsection{Region of validity of the predictions}

The $I$-size distribution $D(\rho)$ and the $\iai$ interaction $\Omega\left(
 U,\frac{ R^2}{\rho\overline{\rho}}, \frac{\overline{\rho}}{\rho} \right)$
form a crucial link between deep-inelastic scattering and 
lattice observables in QCD vacuum\cite{rs-lat}. 

Lattice simulations, on the other hand, provide independent,
non-perturbative information on the 
{\it actual} range of validity of the form predicted from $I$-perturbation
theory for these important functions
of $\rho$ and $R/\rho$, respectively. The one-to-one saddle-point
correspondence (\ref{saddle}) of  the (effective) collective
$I$-coordinates ($\rho^\ast,R^\ast/\rho^\ast$) to $(Q^\prime, \xpr)$
may then be exploited to arrive   
at a ``fiducial'' $(Q^\prime, \xpr)$ region for our predictions in
DIS. Let us briefly summarize the results of this strategy\cite{rs-lat}.
 
We have used the recent high-quality lattice
data\cite{ukqcd,rs-lat} for quenched QCD ($n_f=0$) by the UKQCD
collaboration together with the careful, non-perturbative lattice
determination of the respective QCD $\Lambda$-parameter, 
$\Lambda^{(n_f=0)}_{\overline{\rm MS}}= (238{\pm  19}) {\rm \ MeV}$,
by the ALPHA collaboration\cite{alpha}.  
\begin{figure}
\begin{center}
\parbox{6cm}{\vspace{0.5ex}\includegraphics*[width=6cm]{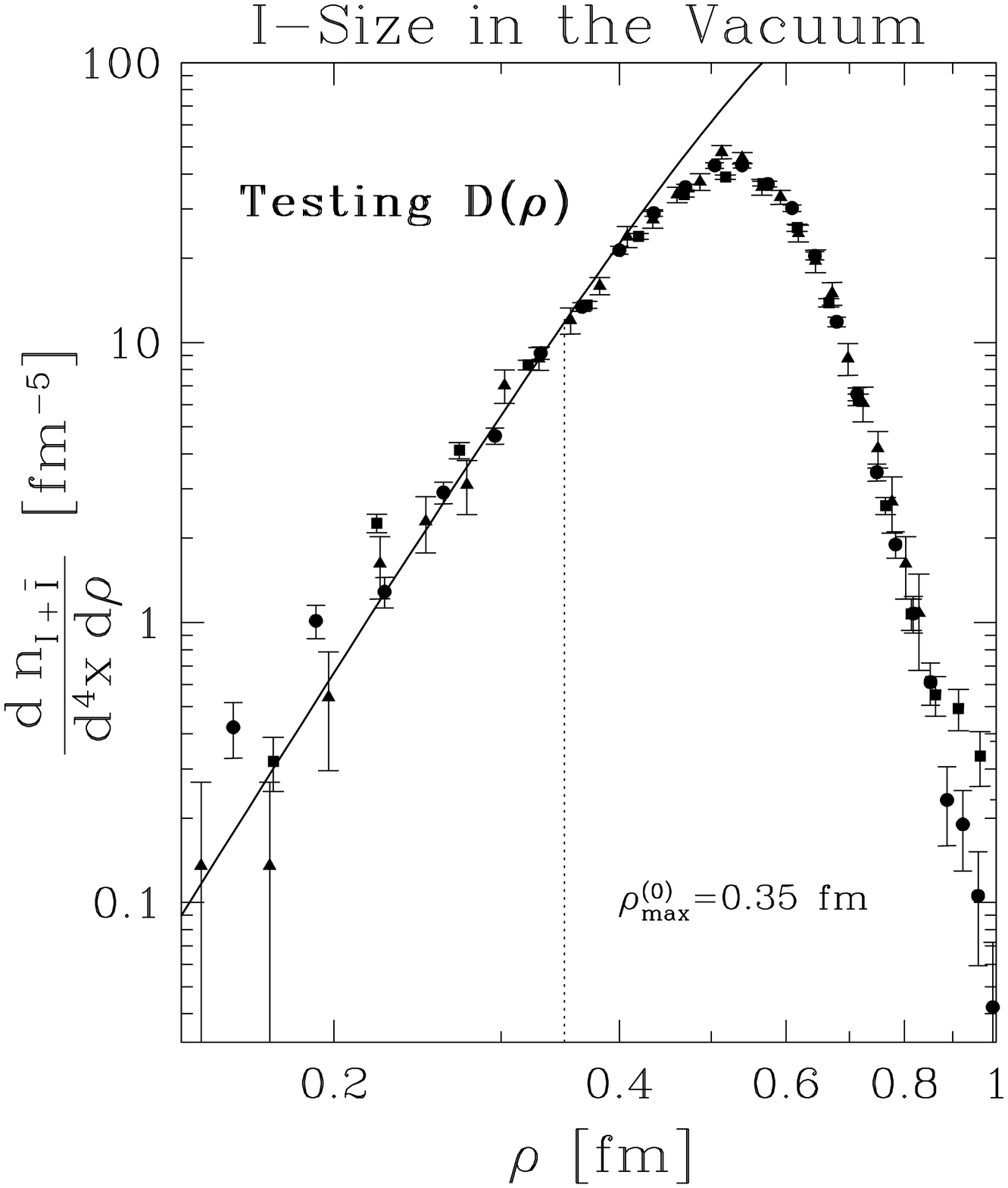}\vspace{0.5cm}}\hspace{0.8cm}\parbox{6cm}{\includegraphics*[width=6cm]{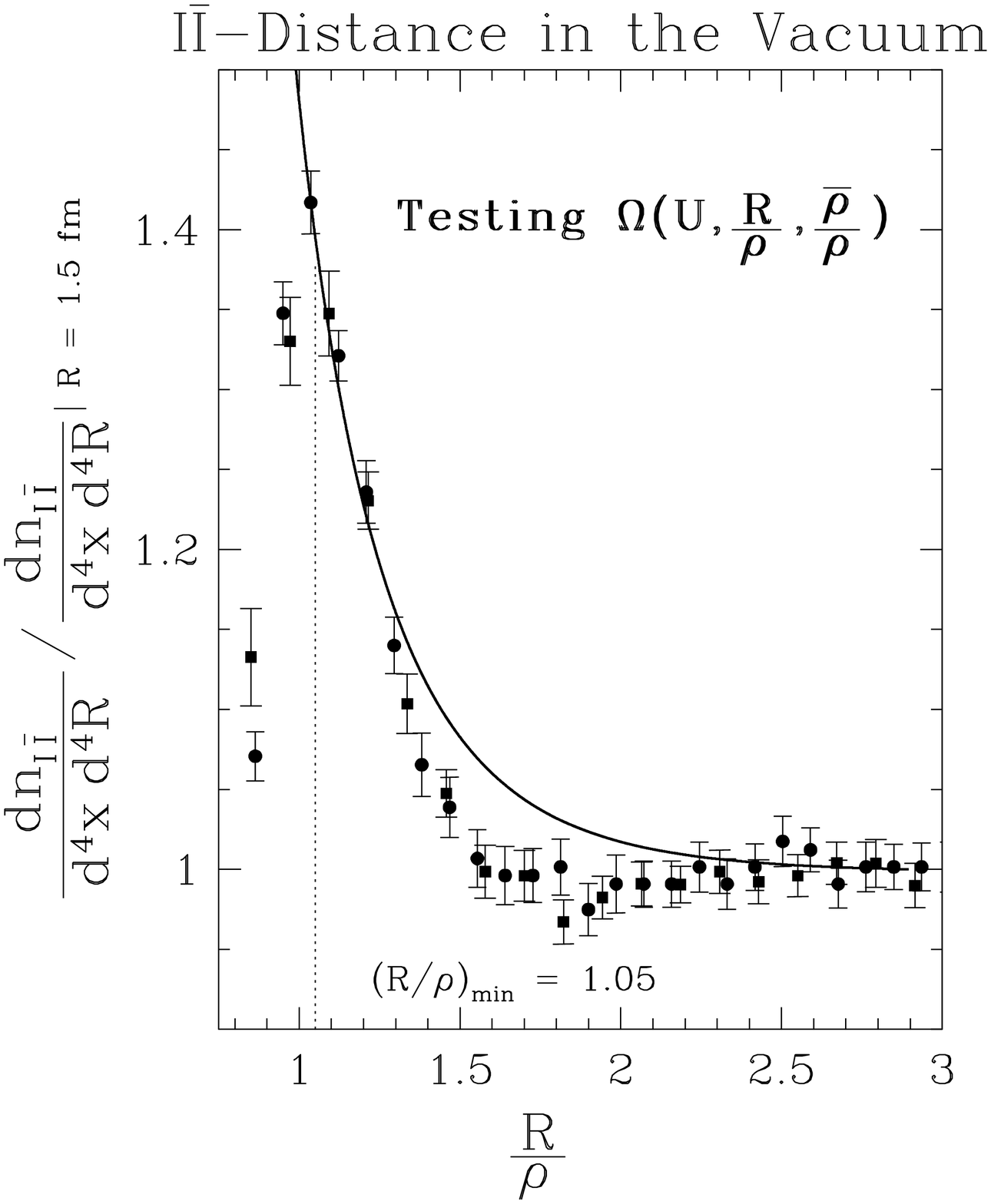}}
\end{center}
\caption[dum]{\label{lattice} Illustration of the agreement
of recent high-quality lattice data~\cite{ukqcd,rs-lat} for the
instanton-size distribution (left) and the normalized $I\overline{I}$-distance
distribution (right) with the predictions from  
instanton-perturbation theory~\cite{rs-lat} for $\rho\lwig
0.35$ fm and $R/\rho\gwig 1.05$, respectively. 
$\alpha_{\rm \overline{MS}}^{\rm 3-loop}$ with $\Lambda_{{\rm
\overline{MS}}}^{(n_f=0)}$ from ALPHA~\cite{alpha} was used.}
\end{figure}
The results of an essentially
parameter-free comparison of the continuum limit\cite{rs-lat} for the
simulated $(I+\overline{I})$-size and the 
$\iai$-distance distributions with $I$-perturbation theory versus
$\rho$ and $R/\rho$, respectively, is displayed in \Fref{lattice}.
The UKQCD data for the $\iai$-distance distribution provide the first
direct test of the $\iai$ interaction $\Omega\left( U,\frac{
R^2}{\rho\overline{\rho}}, \frac{\overline{\rho}}{\rho} \right)$
via\cite{rs-lat}  
\begin{eqnarray} 
\frac{{\rm d}\,n_{I\overline{I}}}{{\rm d}^4x\, {\rm d}^4 R}_{|\rm UKQCD} 
      \stackrel{?}{\simeq}
      \int\limits_0^\infty d\,\rho\,
      \int\limits_0^\infty d\,\overline{\rho}\,
       D(\rho)\, D(\overline{\rho})
      \int d\,U\,{\rm e}^{-\frac{4\pi}{\alpha_s}\,
      \Omega\left(
      U,\frac{R^2}{\rho\overline{\rho}},\frac{\overline{\rho}}{\rho}
      \right)},   
\end{eqnarray}
and the lattice measurements of $D(\rho)$.

From \Fref{lattice}, $I$-perturbation theory appears to be
quantitatively valid for 
\begin{equation}
\left.\begin{array}{lcl}{\rho\cdot\Lambda^{(n_f=0)}_{\overline{\rm
        MS}}}&{ \lwig}&{ 0.42}\\[1ex]{ R}/{ \rho}&{
        \gwig}&{ 1.05}\\ 
        \end{array}
        \right\}\stackrel{\rm\bf saddle\ point}{\Rightarrow}
        \left\{\begin{array}{lcl}{ Q^\prime}/
        {\Lambda^{(n_f)}_{\overline{\rm MS}}}&{ \gwig}&{ 
        30.8},\\[2ex] 
        { x^\prime}&{ \gwig}&{  0.35},\\\end{array} \right .
\label{fiducial}
\end{equation}
Beyond providing a quantitative estimate for the ``fiducial'' momentum
space region in DIS, the good, parameter-free  agreement of the
lattice data with $I$-perturbation theory is very interesting in its own right.
Uncertainties associated with the inequalities (\ref{fiducial}) are studied in
detail in Ref.\,\cite{rs3}.

\subsection{Characteristic final-state signature}
\begin{figure}[ht]
\begin{center}
\parbox{4.2cm}{\includegraphics*[width=4.2cm]{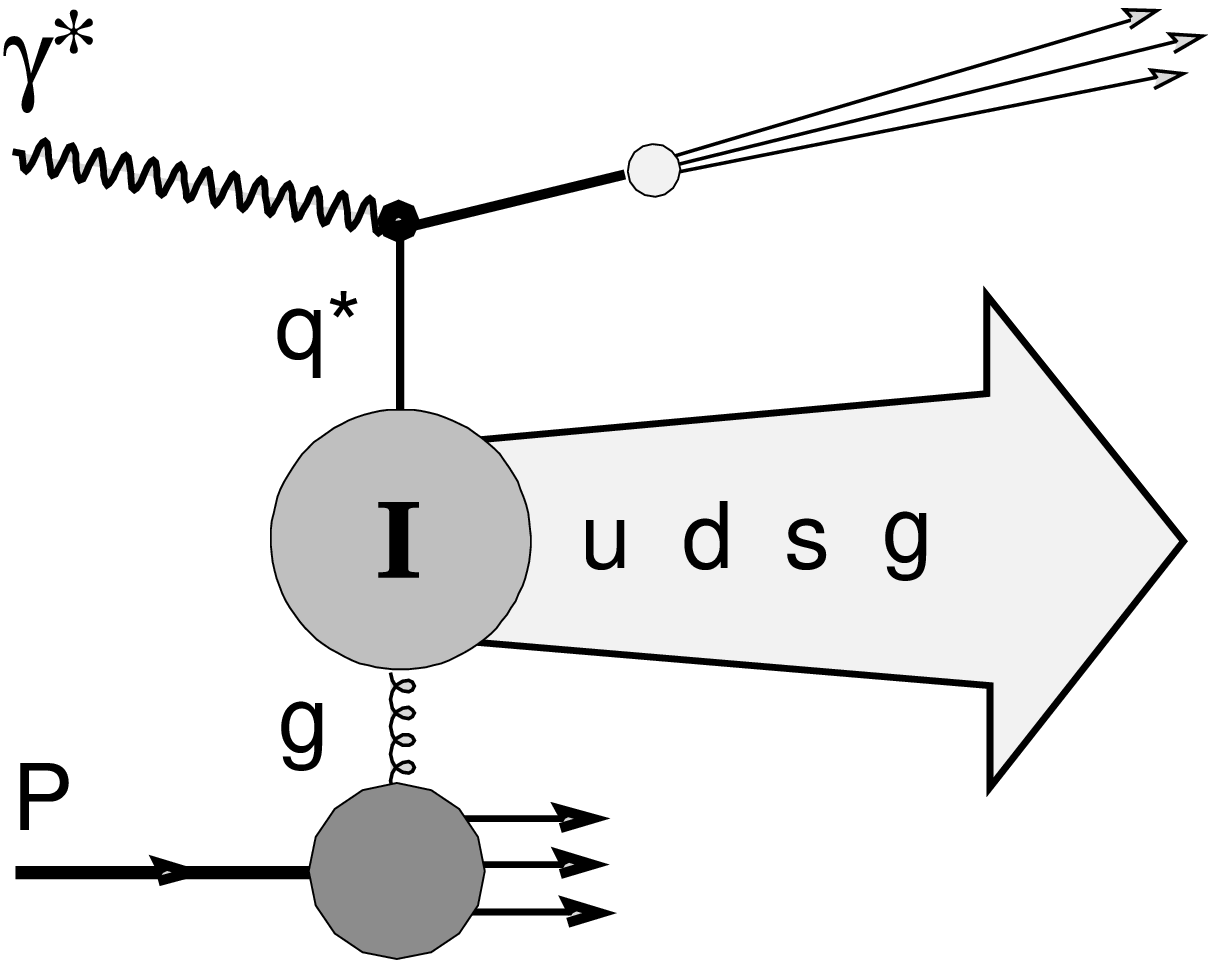}}\hspace{-0.2ex}
\parbox{6.2cm}{\vspace{-3.5ex}\begin{minipage}{6.2cm}{
\small current {\bf jet}\\[0.5cm]
{\bf ``band''}-region:
        ``Fireball'' decaying {\it
        isotropically} (in $I$-rest system) into    
        $n_f\,(q +\overline{q}\,)+\,
        {{\cal O}(\frac{ 1}{\alpha_s})}\,
        g=\mathcal{O}(10)$ partons  
}\end{minipage}}
\parbox{5cm}{\includegraphics*[width=5cm]{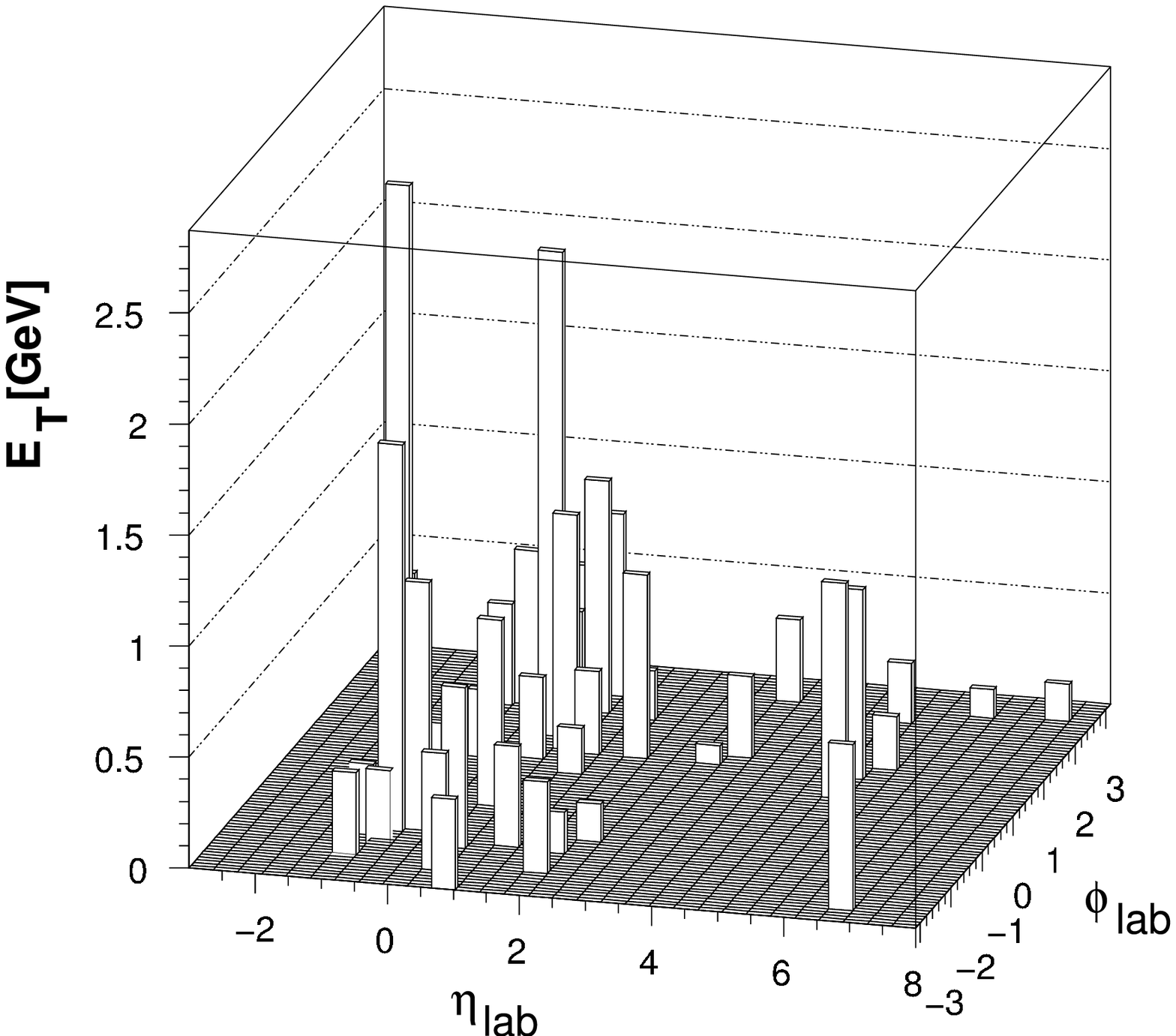}}
\end{center}
\caption[dum]{\label{event}Characteristic signature of $I$-induced
events: {\it One} (current) {\it jet} along with a densely filled {\it
band} of hadrons in the ($\eta,\phi$) plane. Each event has large
hadron multiplicity, large total $E_t$, u-d-s flavor democracy
with 1 $s\overline{s}$-pair/event leading to $K's,\Lambda's\ldots$. An
event from our QCDINS\cite{qcdins} generator (right) illustrates
these features.} 
\end{figure}

The qualitative origin of the characteristic final-state signature of
$I$-induced events is intuitively explained and illustrated in
\Fref{event}.
An indispensable tool for a quantitative investigation of the characteristic
final-state signature and notably for actual experimental searches of
$I$-induced events at HERA is our  Monte-Carlo generator package
QCDINS\cite{qcdins}. 

\subsection{First dedicated search results versus theory}

At the recent DIS2000 and ICHEP2000 conferences, the H1 collaboration
has reported preliminary results of a first
dedicated search for instanton-induced events at HERA\cite{mikocki,h1_ichep}. 
The results presented are quite intriguing  and encouraging, although
far from being conclusive. 
In view of a separate experimental talk on these data at this
meeting\cite{koblitz}, let us briefly summarize and discuss these findings
from a theorist's perspective, while keeping comments on experimental
aspects at a minimum. 

The H1 analysis is based on the strategy\cite{cgrs}
of isolating an ``instanton-enriched'' data sample by means of
suitable cuts to a set of three instanton-sensitive, discriminating
observables. Three different cut-scenarios A), B) and C)
with {\it increasing} instanton-separation power 
$\epsilon_{\rm I}/\epsilon_{\rm nDIS}$ were considered, with
$\epsilon_{\rm I}$ and $\epsilon_{\rm nDIS}$ being the efficiencies
for $I$-events (QCDINS\cite{qcdins} generator) and normal DIS-events (two
generators: CDM=``Color Dipole Model'' (ARIADNE\cite{ariadne}) and
MEPS (RAPGAP\cite{rapgap})), respectively. 

In a phase space region, where a reduction of the normal
DIS (nDIS) background to the percent level is achieved according to
the considered Monte Carlo models, a (statistically) significant
excess of events was found in the H1 data. 
\begin{figure}
\begin{center}
\parbox{4cm}{\includegraphics*[width=4cm]{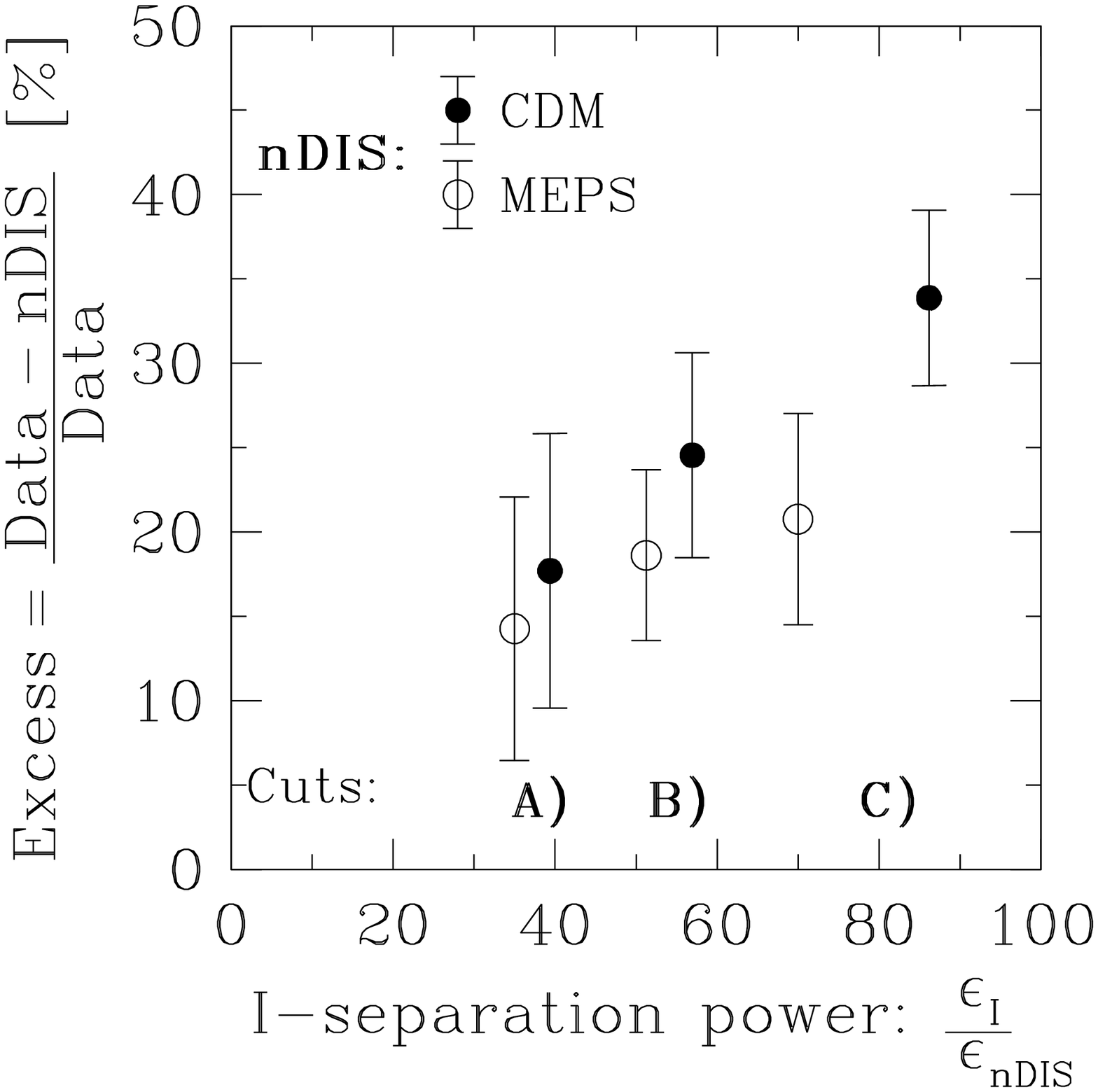}}\hfill
\parbox{11cm}{\includegraphics*[width=11cm]{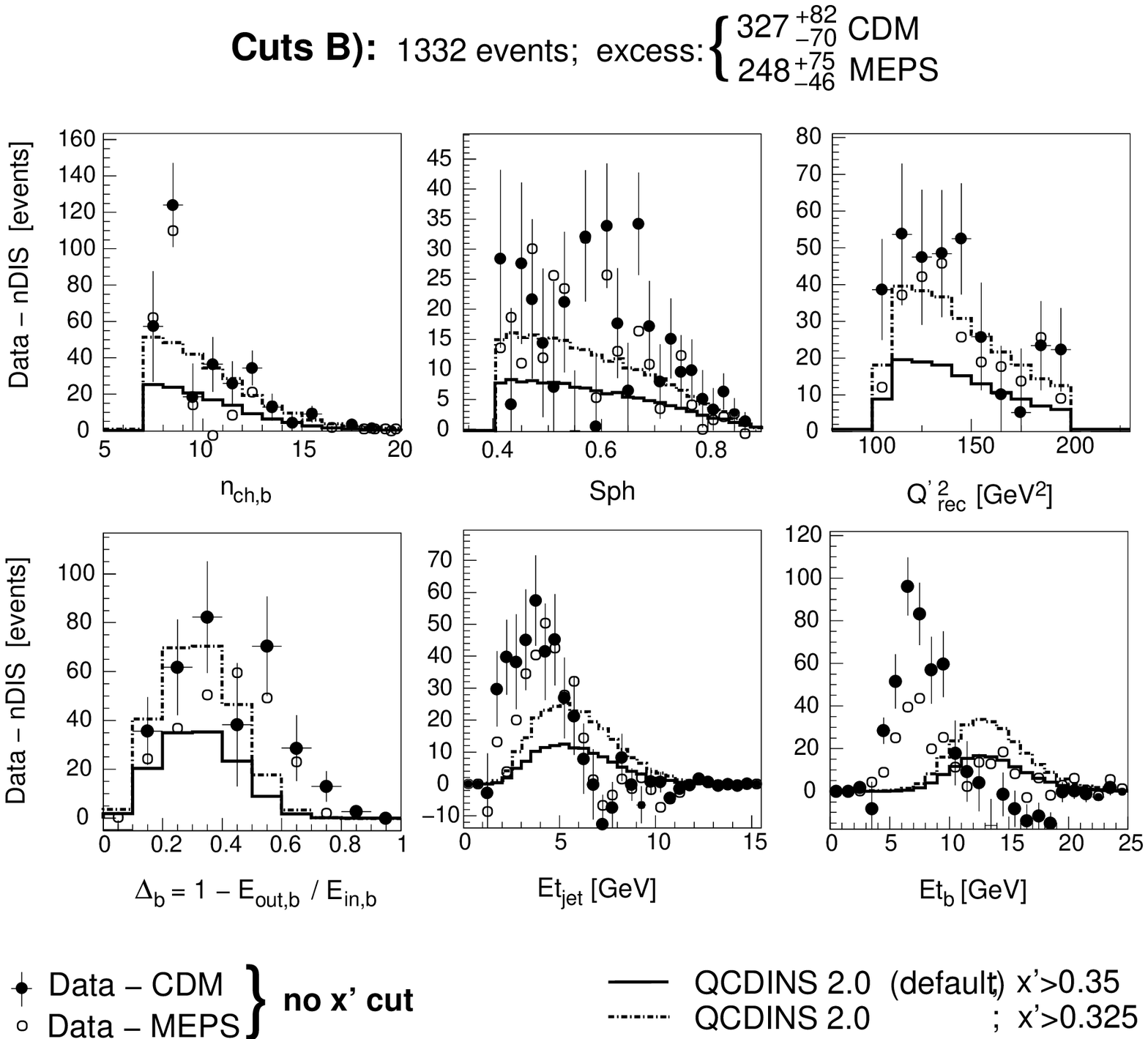}}
\end{center}
\caption[dum]{\label{excess} Summary of preliminary H1 data from a
dedicated search for instanton-induced events\cite{mikocki,h1_ichep}
along with our parameter-free predictions. (left): The observed excess
over normal DIS events according to standard Monte Carlo
generators\cite{ariadne,rapgap} appears to 
increase with the instanton-separation power. (right): Excess in the
six considered observables (\ref{discrim}, \ref{nocut}) compared to our
predictions from QCDINS\cite{qcdins}.    
The size of the instanton-signal is fixed in terms of
$\Lambda^{(5)}_{\overline{\rm MS}} = 219$ MeV (1998 world
average~\cite{pdg98}).} 
\end{figure}
\Fref{excess}\,(left), displays the experimentally observed excess
versus the $I$-separation power, the latter being a purely theoretical quantity
determined from the $I$- and nDIS event generators. It is quite
remarkable that experiment and theory seem to be correlated,
i.\,e. that the excess tends to increase with increasing
$I$-separation power.   

The observed excess along with our original, parameter-free theoretical
predictions\cite{rs2,qcdins} in the six considered
distributions\cite{mikocki,h1_ichep,koblitz} of $I$-sensitive
observables is displayed in \Fref{excess}\,(right).  The
first row shows the three discriminating observables,
\begin{equation}
\mbox{
\begin{tabular}{ll}
$n_{ch,\,b}:$ & number of charged hadrons in the $I$-``band'' region\\
$Sph:$ & sphericity in the rest system of  ``non-jet''-particles\\   
$Q_{\rm rec}^{\prime\,2}:$ & reconstructed virtuality of the $I$-subprocess
quark $q^\prime$  
\end{tabular}
}\label{discrim}
\end{equation}
\noindent
while the second row displays three further observables without
additional cuts applied,
\begin{equation}
\mbox{
\begin{tabular}{ll}
$\Delta_b:$ & $Et$-weighted azimuthal-isotropy  \\   
$Et_{\rm jet}:$ & transverse energy of the current-jet\\ 
$Et_b:$ & total transverse energy in the $I$-``band'' region.
\end{tabular}
}\label{nocut}
\end{equation} 
In the first four of the six observables in
\Fref{excess}, the excess is intriguingly similar in shape and
normalization to our theoretical predictions from QCDINS, although 
its size is partly at a level still comparable to the differences
among the considered DIS event generators.  
While the size of the observed excess in the remaining two observables,
$Et_{\rm b}$ and $Et_{\rm jet}$, is in rough agreement
as well, the peaks of these two experimental distributions appear to be shifted
towards smaller values compared to QCDINS, and the widths
are also considerably narrower. 

Here, some important theoretical comments are in place\cite{rs3}.
One has to take into account that
so far the preliminary H1 data incorporate only part of
the theoretically required cuts: While the $Q^{\prime\,2}\gwig 113$
GeV$^2$ cut (\ref{fiducial}) has been applied to the data, both the 
$x^\prime$-cut (\ref{fiducial}) and notably a further cut  on $Q^2$,
\begin{equation}
Q^2\gwig Q^{\prime\,2}_{\rm min}=113$ GeV$^2,
\label{q2cut}
\end{equation}
(c.\,f. Refs.\,\cite{mrs,qcdins}),  are lacking. 
In Refs.\,\cite{mikocki,h1_ichep,koblitz}, as well as in \Fref{excess}\,(right)
(solid line), these data are compared to the QCDINS output
with {\it active} $x^\prime$-cut (\ref{fiducial}), but
with the default $Q^2$-cut (\ref{q2cut}) switched off to match the data.

The implications of the lacking $ x^\prime$-cut in the data are
presumably not too serious, since QCDINS {\it with} the default
$x^\prime$-cut models to some extent the sharp suppression of
$I$-effects, apparent in the lattice data (c.\,f. \Fref{lattice} (right)) for
$ R/\rho\lwig 1.0-1.05$, i.e. $ x^\prime \lwig 0.3-0.35$.     
Yet, this lacking, experimental cut introduces a substantial
uncertainty  in the predicted  magnitude of the $I$-signal that hopefully
may be eliminated soon. The dash-dotted line in \Fref{excess}\,(right)
illustrates that the overall size of the
$I$-signal strongly depends on the actual value of the
$x^\prime$-cut used in QCDINS. The slightly reduced value of
$x^\prime_{\rm min}=0.325$ in \Fref{excess}\,(right) is certainly
compatible with \Fref{lattice} (right), but improves the agreement with the
observed excess considerably, as compared to the default $x^\prime_{\rm
min}=0.35$.    

Next, consider the effects of the lacking $Q^2$-cut (\ref{q2cut}).  
As a brief reminder\cite{mrs,qcdins}, this cut 
assures in particular the dominance of ``planar'' handbag-type graphs in
$\sigma^{(I)}_{\rm HERA}$ and all final-state observables. 
The non-planar contributions do not share the simple, probabilistic
interpretation of the planar ones, involve instantons with a  size
determined by $1/Q$ rather than $1/Q^\prime$  and are 
both hard to calculate and hard to implement in a Monte Carlo generator.
On account of their known power suppression in $1/Q^2$ and a
cross-check in the simplest case without final-state
gluons\cite{mrs}, they can be safely neglected upon application of
the cut (\ref{q2cut}). Because of these reasons, the non-planar
contributions are {\it not} implemented in the QCDINS event generator,
corresponding  to unreliable QCDINS results for small $Q^2$.

\begin{figure}
\begin{center}
\parbox{6cm}{\includegraphics*[width=6cm,angle=-90]{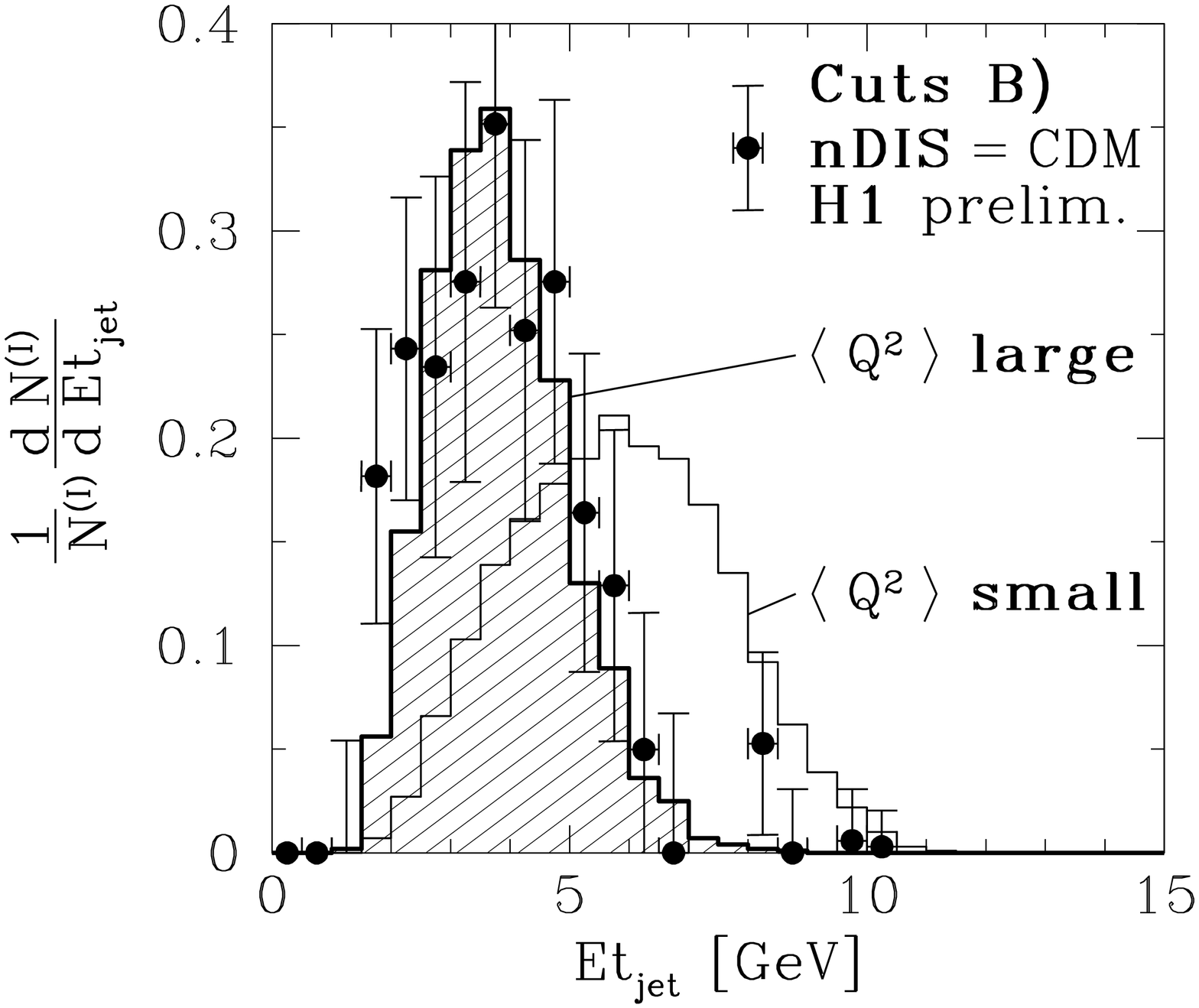}}\hspace{10ex}
\parbox{6cm}{\includegraphics*[width=6cm,angle=-90]{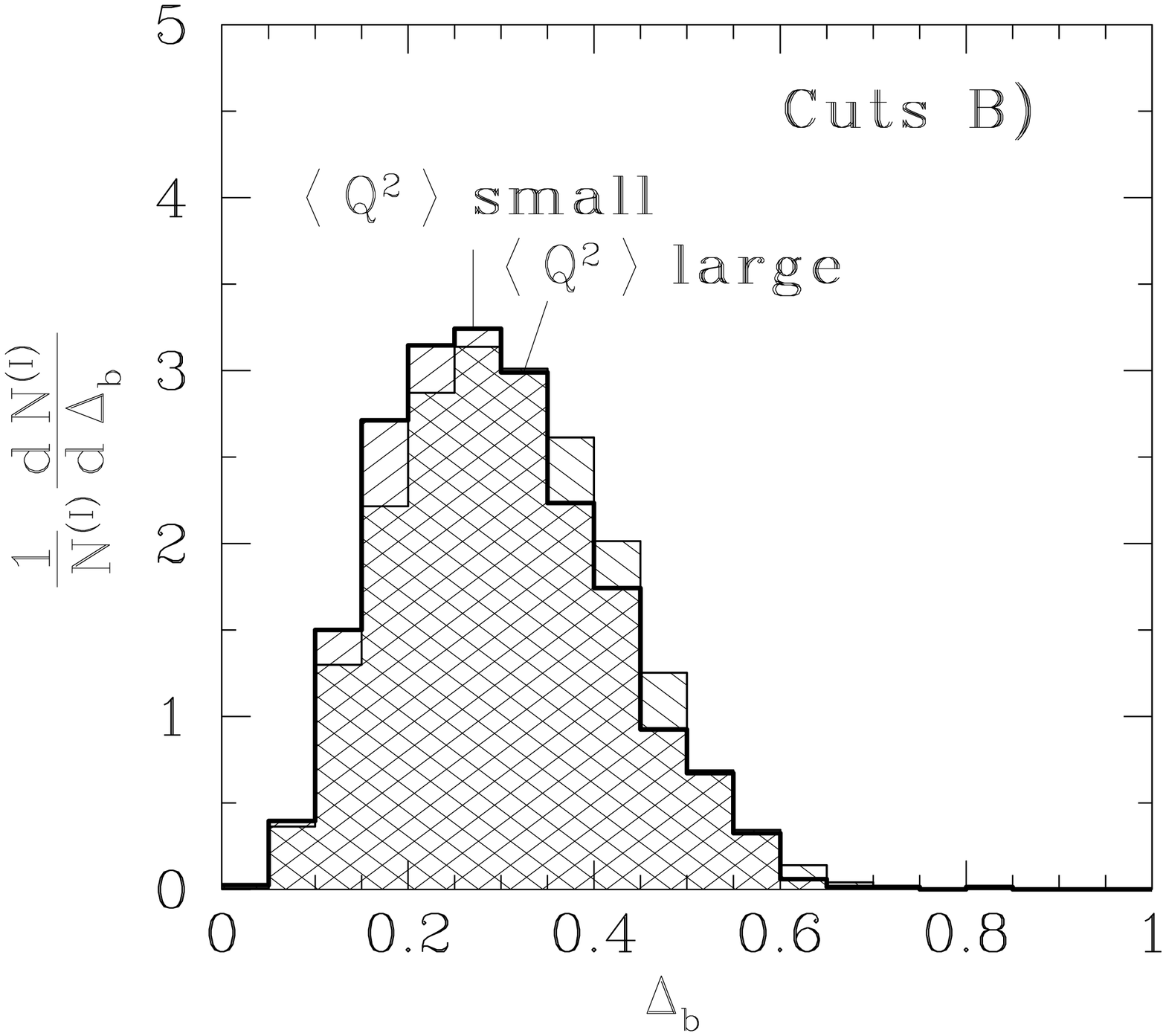}}
\end{center}
\caption[dum]{\label{ptje} Effect of the lacking $Q^2$-cut on the
predicted excess, i.\,e. the $I$-signal.\\
(left): The {\it reliable portion} of generated $I$-events (with high $\langle
Q^2\rangle$) is distributed in $Et_{\rm jet}$ (and in $Et_b$) 
quite like the measured excess (H1
preliminary\cite{mikocki,h1_ichep,koblitz}).  (right): The $Q^2$-cut
has little effect on the {\it shape} of the remaining four observable
distributions. All histograms are from QCDINS\cite{qcdins}.} 
\end{figure}
From a detailed study of the effects of the lacking $Q^2$-cut
(\ref{q2cut}) in Ref.\,\cite{rs3}, the following conclusions have
emerged.
\begin{itemize}
\item The shape of the $Et_{\rm jet}$ and $Et_b$
      distributions for $I$-events are strongly affected by the 
      missing $Q^2$-cut (see e.\,g. $Et_{\rm jet}$ in
      \Fref{ptje}\,(left)), unlike the four remaining observables
      (see e.\,g. $\Delta_b$ in \Fref{ptje}\,(right)). 
\item By cutting out the phase space region where QCDINS is
      unreliable ($Q^2\lwig Q^{\prime\,2}_{\rm min}=113$ GeV$^2$),
      both the resulting $Et_{\rm jet}$ and $Et_b$-peaks are left-shifted  
      and become narrower, in good agreement with the
      shape-normalized H1-excess (see e.\,g. $Et_{\rm jet}$ in
      \Fref{ptje}\,(left)). While further reaching conclusions will
      require the actual implementation of the $Q^2$-cut (\ref{q2cut}) in the
      data, this exercise shows that at least the {\it reliable
      portion} of generated $I$-events (with high $\langle
      Q^2\rangle$) is distributed in $Et_{\rm jet}$ and in $Et_b$
      quite like the observed excess.  
\end{itemize}
Despite an improved overall agreement and understanding via such
considerations, let us close this section with a reminder of the basic
remaining problematics: The observed excess is strongly relying on
Monte Carlo generators for {\it normal} DIS events, and the $I$-signal
is expected where the latter are not too well known/studied \ldots\,.
\section{Larger-size instantons}
This section is devoted to some ongoing attempts towards a better
understanding of the r{\^o}le of larger-size instantons both in the
QCD-vacuum and in high-energy (diffractive) scattering processes. The 
guidance from recent high-quality lattice simulations is indispensable
in this regime, as will become evident next. 
\subsection{A residual conformal inversion symmetry?}
\begin{figure}
\begin{center}
\includegraphics*[height=13cm,angle=-90]{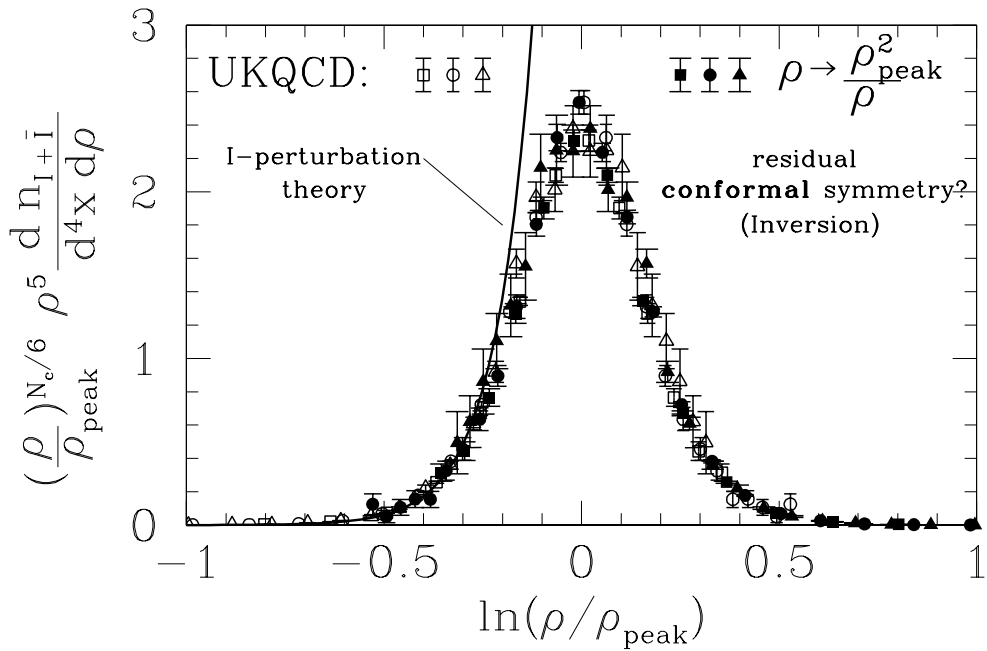}
\end{center}
\caption[dum]{\label{isize} Same UKQCD lattice data\cite{ukqcd,rs-lat} for the
instanton-size distribution as in \Fref{lattice}, displayed however
such as to suggest a virtually perfect {\it inversion symmetry} under
$\rho\Rightarrow \rho_{\rm peak}^2/\rho$ with $\rho_{\rm peak}\approx 0.6$ fm
(open and solid data symbols fit onto one universal, symmetric
curve). The solid line refers to instanton-perturbation theory
analogous to \Fref{lattice}.}   
\end{figure}
Let us start by considering again the high-quality UKQCD lattice
data\cite{ukqcd,rs-lat} for 
the $I$-size distribution ${{\rm d}\,n}/{{\rm d}^4x\, {\rm d}\rho}$. Unlike 
\Fref{lattice}\,(left), in \Fref{isize} the dimensionless quantity
$(\rho/\rho_{\rm peak})^{(N_c/6)}\,\rho^5\,{{\rm d}\,n}/{{\rm d}^4x\, {\rm
d}\rho}$ is now displayed versus $\ln(\rho/\rho_{\rm peak})$, with
$\rho_{\rm peak}= 0.605$ fm being the (empirical) peak position.

The lattice data in \Fref{isize} first of all illustrate the
striking fact that larger-size instantons are dramatically suppressed
in comparison to (naive) expectations from $I$-perturbation theory
(solid line). The peak position, $\rho_{\rm peak}$, may be viewed as
an important length scale, characterizing the fairly rapid
breakdown of usual $I$-perturbation theory. It is clearly important,
to ask what kind of underlying physics can give rise to such a rapid
and dramatic change. The focus here will be on an intriguing,  possible
explanation\cite{conformal} in terms of a residual conformal
{\it inversion symmetry}.    

The reason for displaying the lattice data
in \Fref{isize} versus $\ln(\rho/\rho_{\rm peak})$ was to make the virtually   
perfect 
\begin{equation}
\mbox{\rm {\it inversion} symmetry:\ } \rho \Leftrightarrow \frac{\rho_{\rm
peak}^2}{\rho}
\label{inversion}
\end{equation}
self-evident in the lattice data. Both the open data symbols, referring to
the original data points, and the solid ones, involving inverted
arguments according to (\ref{inversion}), seem to beautifully fit onto one
universal, symmetric curve.   

The possibility of such an inversion symmetry is particularly
appealing, since it may well be a ``relict'' from the well-known { \it
conformal invariance} of the whole $I$-sector at the classical
level\cite{jackiw}.  If true,
such a symmetry would allow to access the non-perturbative regime of
large-size instantons (yet with {\it small} $\rho^\prime=\rho_{\rm peak}^2/\rho$) in
terms of $I$-perturbation theory for instantons with {\it small} $\rho$.   
It may well have also intriguing consequences  
beyond instanton physics for QCD in general, as we shall briefly speculate
further below. 

The conformal symmetry group comprises\footnote{In Euclidean space, the analog of
Poincar{\'e} invariance is Euclidean invariance. Note moreover, that
the inversion is not continuously connected to the identity. In order
to discuss conformal transformations infinitesimally, one usually
considers instead the so-called  special conformal transformations
that involve an inversion followed by a translation and another inversion.} 
besides the inhomogeneous Lorentz group (Poincar{\'e}
invariance), scale transformations, $x^\prime_\mu \Rightarrow
x_\mu/\lambda$, and the coordinate {\it inversion:}, 
\begin{equation}
x_\mu \ \Rightarrow\  x^\prime_\mu=\frac{\rho_0^2}{ x^2}\,x_\mu.
\label{coordinv}
\end{equation}  
The invariance under scale transformations is well-known to be broken
at the quantum level via regularization/renormalization.    
A detailed and notably more rigorous discussion about the validity of
the conformal inversion at the quantum level ($I$-size distribution!),
would certainly  lead beyond the scope of this 
talk and may be found elsewhere\cite{conformal}. Yet, it may  
be instructive to sketch a few simple arguments from
Ref.\,\cite{conformal}. First of all, let us ask at the classical
level, why a coordinate inversion (\ref{coordinv}) 
indeed implies an inversion (\ref{inversion}) of the instanton
size. For simplicity, let us consider a pure SU(2) Yang-Mills theory
(no fermions). Starting from the familiar expression for the
vector potential of an instanton in singular gauge (gauge
coupling g), 
\begin{equation}
A^{(I)\,a}_\mu
(x;\,\rho)=\frac{2}{g}\,\frac{\rho^2}{x^2}\,\frac{\overline{\eta}_{a
\mu \nu}\,x^\nu}{x^2+\rho^2},
\end{equation}
involving the 't Hooft coefficients\cite{th} $\overline{\eta}_{a \mu
\nu}$, one straightforwardly finds for the transformed vector field 
$A^{\prime\,\,(I)\,a}_\mu (x;\,\rho)$ after a conformal coordinate
inversion (\ref{coordinv}),  
\begin{eqnarray}
\fl A^{(I\,)\,a}_\mu (x;\,\rho)\ \Rightarrow\
A^{\prime\,\,(I)\,a}_\mu (x;\,\rho)= A^{(I)\,a}_\nu
(x^\prime;\,\rho)\,\frac{{\rm d}x^{\prime\,\nu}}{{\rm d}x^\mu}  
=A^{(\overline{I}\,)\,a}_{\mu}(x;\frac{\rho_0^2}{\rho})_{\mid \,{\rm
regular\,gauge}}\,.   
\end{eqnarray}
Apparently, the inversion (\ref{coordinv}) transforms the vector potential
in singular gauge of an instanton with size $\rho$ into the vector
potential in regular gauge of an {\it anti}-instanton with size
$\rho_0^2/\rho$ ! Using the conformal transformation law of the 
field-strengh tensor,
\begin{equation}    
G^a_{\mu\nu}(x) \Rightarrow G^{\,\prime\,a}_{\mu\nu}(x)=G^a_{\alpha\beta}(x^\prime)\frac{{\rm d}x^{\prime\,\alpha}}{{\rm d}x^\mu}\frac{{\rm d}x^{\prime\,\beta}}{{\rm d}x^\nu},
\end{equation} 
on arrives after some calculation at the following (gauge-independent)
transformation of the (anti-) instanton contribution to the Lagrange density,
\begin{equation}
\mathcal{L}^{(I)}(x,\rho)=\frac{1}{4}\,G^{(I)\,a}_{\mu\nu}(x,\,\rho)
\,G^{(I)\,a\,\mu\nu}(x,\,\rho)\Rightarrow
\mathcal{L}^{\,\prime\,(I)}(x,\rho)=\mathcal{L}^{(I)}(x,\frac{\rho_0^2}{\rho}),
\end{equation} 
which illustrates the $I$-size inversion (\ref{inversion}) of interest
as resulting from the coordinate inversion (\ref{coordinv}). Upon
integration over space-time one then explicitly checks the invariance of the
$I$-action $S^{(I)}$ under coordinate inversion due to its
independence of the $I$-size,
\begin{equation}
S^{(I)}\equiv\int d^4x  \mathcal{L}^{(I)}(x,\rho) = \int d^4x
\mathcal{L}^{(I)}(x,\frac{\rho_0^2}{\rho}) = \int d^4x
\mathcal{L}^{\,\prime\,(I)}(x,\rho)=\frac{8\pi^2}{g^2}.
\end{equation} 
The essential step, however, is to reconsider\cite{conformal} the
derivation of the (1-loop) vacuum-to-vacuum
(tunnelling) amplitude\cite{th,bernard} about a single instanton that directly
determinines the leading expression for the $I$-size distribution. 
The task is to compare the result for instantons with {\it small}
$\rho$ to that for instantons with {\it small} $\rho^\prime=\rho_{\rm peak}^2/\rho$.
It turns out that the (dominating) {\it zero-mode} contribution indeed
respects the 
inversion symmetry (\ref{inversion}), while a non-invariant piece due to
the various non-zero mode determinants has been studied long ago\cite{yoneya}, 
may be isolated and then divided out in form of the factor
$(\rho/\rho_{\rm peak})^{(N_c/6)}$ in \Fref{isize}. 

An important challenge in this approach relies in a better
understanding of the significance of the inversion scale $\rho_{\rm
peak}$. While unbroken scale invariance would (nonsensically) make
{\it any value} of $\rho_{\rm peak}$ physically equivalent, its 
breaking signalled by the non-vanishing trace of the energy-momentum
tensor\cite{anomaly}, $\theta^\mu_\mu \propto
-\langle0\mid\frac{\alpha_s}{\pi}\,G_{\mu\nu}^{a\,2}\mid 
0\rangle$, suggests $\rho_{\rm peak} \sim \langle 0\mid\frac{\alpha_s}{\pi}\,
G_{\mu\nu}^{a\,2}\mid 0\rangle^{-1/4}$. 

Let us close with pointing out an intriguing possible consequence of such an
inversion symmetry of the $I$-size distribution, that affects
$\alpha_s$ and thus QCD in general. Let us follow Ref.\,\cite{rs-lat} and 
define a (non-perturbative) ``$ I$-scheme'' for
$\alpha_s(\mu_r)$, after identifying the renormalization scale
$\mu_r$ as $\mu_r=\frac{s_I}{\rho}$ with $s_I=\mathcal{O}(1)$, by the
requirement that the familiar perturbative expression of
$\rho^5\,{{\rm d}\,n}/{{\rm d}^4x\, {\rm d}\rho}\,[\alpha_s,s_I]$ be
valid for all $\alpha_s(\frac{s_I}{\rho})$. Surprisingly, the form of
$\alpha_s(\frac{s_I}{\rho})$, implicitly defined by this prescription
and directly extracted from a comparison with the UKQCD
data\cite{ukqcd,rs-lat}, showed a 
Cornell form $\alpha_s \approx \frac{3}{4}\, \sigma\,\rho^2+\ldots$ for
$\rho\gwig\rho_{\rm peak}$ with string tension
$\sqrt{\sigma}\approx 472$ MeV, while beautifully agreeing with the
3-loop perturbative form of $\alpha_{\overline{\rm MS}}$ for
$\rho\lwig \rho_{\rm peak}$. Taking here for the sake of
simplicity the leading 1-loop expression for the $I$-size distribution\cite{th,bernard}, the inversion symmetry (\ref{inversion}) of the quantity
in \Fref{isize} implies
\begin{eqnarray}
\fl \left(\frac{\rho}{\rho_{\rm peak}}\right)^{(N_c/6)}\,\rho^5\,\frac{{\rm
d}\,n_I}{{\rm d}^4x\, {\rm d}\rho} 
&\equiv&{\rm const.}\left(\frac{\rho}{\rho_{\rm
peak}}\right)^{(N_c/6)}\,\left(\frac{2\pi}{\alpha_s(\frac{s_I}{\rho})}\right)^{2
N_c}\,\exp(-\frac{2\pi}{\alpha_s(\frac{s_I}{\rho})})\nonumber\label{sym}\\
&=&{\rm const.}\left(\frac{\rho_{\rm peak}}{\rho}\right)^{(N_c/6)}\,\left(\frac{2\pi}{\alpha_s(\frac{s_I\,\rho}{\rho_{\rm peak}^2})}\right)^{2
N_c}\,\exp(-\frac{2\pi}{\alpha_s(\frac{s_I\, \rho}{\rho_{\rm
peak}^2})}). 
\end{eqnarray}
We may explicitly solve \Eref{sym} for $\alpha_s(\frac{s_I}{\rho})$ in terms of
$\alpha_s(\frac{s_I\, \rho}{\rho_{\rm peak}^2})$, which involves the
Lambert-W function, $W(x)\,\exp(W(x))=x$. By using the leading,
asymptotically free form, for $\alpha(\frac{s_I\,{\rho}}{\rho_{\rm
peak}^2})\stackrel{ \rho \ {\rm large}}{\approx}
\frac{2\pi}{\beta_0\,\log(\frac{s_I\,{\rho}}{\rho_{\rm
peak}^2\,\Lambda})}+\ldots$, the inversion symmetry (\ref{inversion})
then analytically leads again to a Cornell form  
\begin{equation}
\fl \alpha_s(\frac{s_I}{\rho})\stackrel{\rho \mbox{\rm\
large}}{\approx}-\frac{\pi}{N_c\,{\rm W}(\frac{11}{6}\left(\frac{\rho_{\rm peak}}{\rho}\right)^2  
\,\ln(\frac{\rho_{\rm peak}}{\rho}))}\approx\frac{2\pi}{\beta_0}\,\frac{1}{\ln(\frac{\rho}{\rho_{\rm
peak}})}\,\left(\frac{\rho}{\rho_{\rm peak}}\right)^2 +\mathcal{O}(1),
\label{cornell}
\end{equation}
signalling confinement. More specifically, one may write down a
simple and {\it exact} closed solution of \Eref{sym},
($\beta_0=\frac{11}{3}\,N_c$), 
\begin{equation}
\alpha_s(\frac{s_I}{\rho})=\frac{2\,\pi}{\beta_0}\,\frac{(1-\left( 
\frac{\rho}{\rho_{\rm peak}}\right)^2)}{\ln(\frac{\rho_{\rm peak}}{\rho})},
\ \mbox{\rm with\ } \rho_{\rm peak}=\frac{s_I}{\Lambda},
\label{nesterenko}
\end{equation}
which apparently now has ``inherited'' the inversion symmetry
(\ref{inversion}),  
\begin{equation}
\left(\frac{\rho}{\rho_{\rm peak}}\right)\,\alpha_s(\Lambda\,\frac{\rho}{\rho_{\rm peak}})=\left(\frac{\rho_{\rm
peak}}{\rho}\right)\, \alpha_s(\Lambda\,\frac{\rho_{\rm peak}}{\rho}),  
\end{equation}
has {\it no Landau pole}, the correct  asymptotic freedom form for
$\rho\Rightarrow 0$ as well as a Cornell form (\ref{cornell}) for large $\rho$.
Amazingly, this (1-loop) form (\ref{nesterenko}) of $\alpha_s$ exists
already in the 
literature\cite{nesterenko}, but originated from an entirely different
reasoning.  It appeared as the appropriate (1-loop) running coupling without a
Landau pole in sort of a renormalization-group improved variant of
Shirkov's ``analytic perturbation theory''\cite{shirkov}. 
\subsection{Instanton contribution to the color-dipole picture}
Larger-size instantons may well play an important r{\^o}le in high-energy
scattering processes\cite{shuryak} and even be the driving
configurations for diffractive scattering\cite{pomeron-ins},
i.\,e. the (soft) $\funp$omeron. 
\begin{figure}
\begin{center}
\includegraphics*[width=8cm]{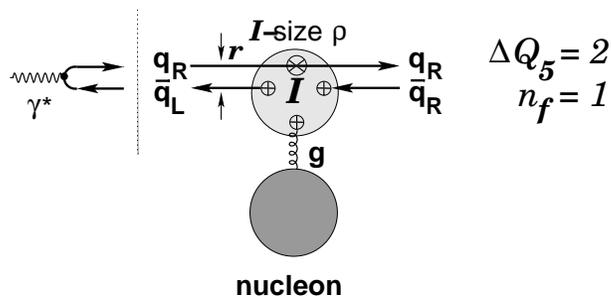}
\end{center}
\caption[dum]{\label{dipole} Illustration of the one-instanton
contribution to the color-dipole picture\cite{dipole} for
the simplest instanton-induced process with $n_f=1$ and no final-state
gluons\cite{mrs}.}  
\end{figure}
As a well-defined start towards these important, but complicated
issues, let us consider (initially at least) the DIS regime where
$I$-perturbation theory holds. In this region, we then have
recast\cite{dipole-ins} our published calculation of the simplest
relevant $I$-induced process\cite{mrs}, 
$\gamma^\ast\,g\stackrel{(I)}{\Rightarrow} q_R\,\overline{q}_R$, with
one massless quark flavour into the language
of the popular color-dipole picture\cite{dipole} (c.\,f. \Fref{dipole}). 
The intuitive content of the latter is that at high energies, in the 
proton's rest frame, the virtual photon fluctuates predominantly into a
$q\overline{q}$ {\it dipole} a long distance upstream of the target
proton. The large difference of the $q\overline{q}$-formation and
$(q\overline{q})$-$P$ interaction times then generically gives rise to
the familiar factorized expression of the inclusive photon-proton
cross sections 
\begin{equation}
\sigma^{\rm L\,T}_{\gamma^\ast\,P}=\int dz\, d^2\vec{r} \mid \Psi^{\rm
L\,T}_\gamma\mid^2\,\sigma_{\rm dipole}.
\label{dipole-cross}
\end{equation} 
in terms of the modulus squared of the photon's (light-cone)
wavefunction $\Psi^{\rm L\,T}_\gamma$ and  
the $(q\overline{q})$-$P$ dipole cross section $\sigma_{\rm dipole}$.
The important variables in \Eref{dipole-cross} are the transverse
$(q\overline{q})$-size $\vec{r}$ and the photon momentum fraction $z$
carried by the quark.  

This dipole picture represents a convenient framework for discussing
the transition from hard to soft physics (diffraction), with expectations
\begin{eqnarray}
\begin{array}{lclcl}
\fl\mbox{\rm ``hard'' (pQCD)\cite{dipole-pqcd}}&:&\ {\it small}\
\vec{r}^2\sim\frac{1}{Q^2},\  \sigma_{\rm dipole}&=&\alpha_s\,\mathcal{O}(\vec{r}^2),\mbox{\rm\ ``color transparency''}\\
\mbox{\rm ``soft''}&:&\sqrt{\vec{r}^{\,2}}\gwig 0.5 {\rm\ fm},\hspace{2ex}\, \sigma_{\rm
dipole}&\approx& {\rm constant},\mbox{\rm\ ``hadron-like''.} 
\end{array}
\label{transition}
\end{eqnarray}
In the large $Q^2$ regime,  where $I$-perturbation holds,
the $I$-contribution (\Fref{dipole}) may indeed be cast approximately
into a form (\ref{dipole-cross}) and be compared to the familiar pQCD
expression\cite{dipole-pqcd} (\ref{transition}). For reasons of space,
let us just emphasize an emerging, striking feature: Unlike the pQCD
expression\cite{dipole-pqcd} for $\sigma_{\rm dipole}$, 
the required integrations over the $I\ (\overline{I})$-sizes now bring
unavoidably as further length scale the effective $(I$-${\rm\,size})^2$,
$\langle \rho\overline{\rho}(Q^2,z)\rangle\propto \int\int
d\rho\,d\overline{\rho}\,\rho^5D(\rho)\,\overline{\rho}^5\,
D(\overline{\rho})\{\ldots\}$, into the nominator  that competes on
dimensional grounds with the
square of the transverse $(q\overline{q})$-size $\vec{r}^{\,2}$ for
$\vec{r}^{\,2}\Rightarrow 0$. Since for a range of moderately high $Q^2$, the
effective $I$-size is dominated, however, by the peak-position
$\rho_{\rm peak}\approx 0.6$ fm of $\rho^5\,{{\rm d}\,n}/{{\rm d}^4x\,
{\rm d}\rho}$ (c.\,f. \Fref{isize}), the instanton in the background
of the $(q\overline{q})$-pair  acts rather like an object of {\it
hadronic} size and a roughly constant $I$-induced dipole cross section
results,  
\begin{equation}
\sigma^{(I)}_{\rm dipole}\propto\frac{1}{\alpha_s}\,\langle
\rho^2\rangle\, \approx {\rm constant},
\end{equation}
even for quite small $(q\overline{q})$-size $\vec{r}^2$. A detailed
analysis of the delicate interplay of these two crucial length scales
may be found in Ref.\,\cite{dipole-ins}. Thus, it seems that due to
this precocious lack of ``color transparency'', 
indeed, instanton  configurations may well play a distinguished r{\^o}le in
building up diffractive high-energy scattering. 
\subsection*{Acknowledgements}
All results on ``small instantons and deep-inelastic scattering''
(Section 2) have been obtained in long-term collaboration  with
Andreas Ringwald. Furthermore, I owe much insight about larger-size
instantons to many discussions with Hans Joos.

\end{document}